# Loss Comparison of Electric Vehicle Fuel Cell Integration Methods

Yuheng Wang, Mehanathan Pathmanathan and Peter W. Lehn

Department of Electrical and Computer Engineering, University of Toronto

(Preprint)

**ABSTRACT**

This study analyzes and compares the drivetrain losses of two methods of fuel cell integration in electric vehicle drivetrains. The first is a conventional (boosted) two-stage system while the second is a dual inverter-based solution. Each source of drivetrain losses is described by mathematical equations and the impact of higher order harmonics are observed through circuit model simulation. The dual inverter system achieves an overall energy efficiency improvement of 5.27% and 10.13% for highway and urban driving compared to the conventional method. The use of lower voltage rated power modules and lower switching frequency in the dual inverter system has significantly reduced the switching losses and improved the driving efficiency.

1. **INTRODUCTION**

Hydrogen fuel cells (FCs) are gaining popularity as alternative electric vehicle (EV) power sources to lithium batteries [1, 2]. FCs have better range, lower cost and less environmental impact in the production process. Despite various economic benefits, the integration of FCs on EVs requires additional effort. FCs can only provide unidirectional power flow (a FC cannot absorb power), and will shutdown if the power production falls below a certain minimum value [3]. Additionally, the output voltage of FC stacks is typically lower than that of battery packs, and its rate of change need to be limited to reduce the risk of damage to the FC [4, 5].

The conventional method to integrate FCs into EV drivetrains is using a boost converter to step up the lower voltage of a FC stack to the higher voltage of an EV battery pack (upper image of Fig. 1) [3]. However, such converters require a magnetic energy storage stage which adds undesirable mass and volume to an EV [1].

The dual inverter based electric vehicle (EV) drivetrain has been proposed as an alternative to conventional two-stage drivetrains [6]. By implementing a well-designed power sharing algorithm, FC can be directly integrated in a dual inverter drive as one of the energy sources and eliminate the DC/DC converter (lower image of Fig. 1). The power sharing algorithm will generate power reference from the vehicle's acceleration profile to operate FC under limitations. Similar power sharing algorithms have been used to coordinate power flow in battery-supercapacitor dual inverter configurations [7].

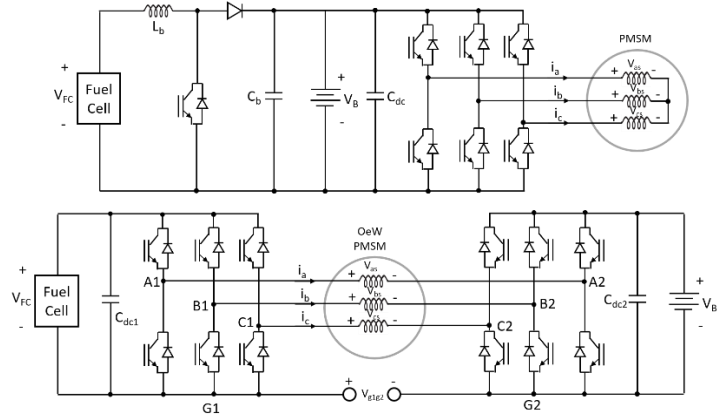

Fig. 1. Conventional (top) and Dual-inverter (bottom) methods of fuel cell integration in electrical vehicles

This paper compares the power loss of a FC-battery hybrid EV drivetrain implemented using a conventional two-stage inverter drive or a dual inverter drive. Power losses of both systems are computed analytically and verified by simulations. The range extension anticipated for highway and driving cycles are evaluated to assess the improvements provided by the proposed dual inverter system over the conventional two-stage converter.

2. **LOSS MODELS**

   A. Traction Motor Model

A three-phase permanent magnet synchronous motor (PMSM) can be represented using the two-axis model shown in Fig. 2.

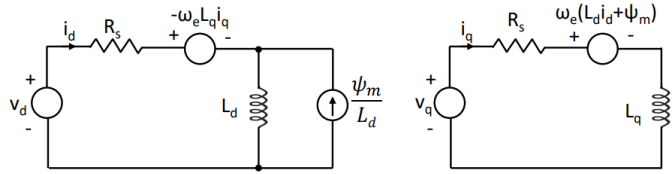

Fig. 2. Permanent magnet synchronous motor dq0 equivalent circuit model

Applying Kirchoff's Voltage Law to this model will give the following equations:

$$v_d = R_s i_d + L_d \frac{di_d}{dt} - \omega_e L_q i_q \tag{1}$$

$$v_q = R_s i_q + L_q \frac{di_q}{dt} + \omega_e (L_d i_d + \psi_m) \tag{2}$$

where $\omega_e$ is the electrical frequency of the motor and $\psi_m$ is its permanent magnet rotor flux linkage.

The electrical power of a PMSM can be expressed in the *dq* frame as:

$$P = \frac{3}{2}(v_d i_d + v_q i_q) \qquad (3)$$

B. Power Sharing Algorithm

A power sharing algorithm is implemented on the dual-inverter system to i) maintain the FC power above the minimum threshold to avoid shutdown, ii) ensure the FC power flow is unidirectional, and iii) make sure the FC power has a slow rate of change. The battery pack supplies the high frequency component of the drivetrain power and absorbs the power from regenerative breaking. Additionally, the low output voltage of the fuel cell under high loads is mitigated by the dual inverter drives ability to generate an overall motor voltage vector which is a composite of the voltage vectors produced by the FC and battery inverters (Fig. 3):

$$v_d = v_{dFC} + v_{dBat} \qquad (4)$$

$$v_q = v_{qFC} + v_{qBat} \qquad (5)$$

Where $v_{dFC}, v_{qFC}, v_{dBat}$ and $v_{qBat}$ are the $d$ and $q$ axes voltages produced by the FC and battery inverters respectively. Careful selection of these values will allow the FC output power to track a reference value. As suggested by [3], this fuel cell power reference is generated as a low pass filtered value of the electrical power required by the PMSM.

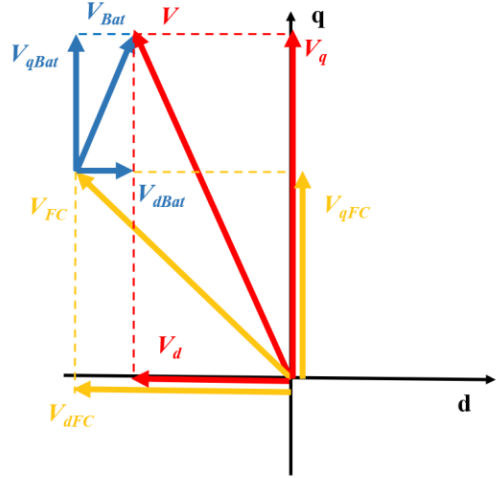

Fig. 3. Vector diagram of dual inverter drive

C. Drivetrain Power Losses

For each power module in the inverter, its conduction and switching losses are functions of the on-state current $I_s$. From [8], the conduction losses of IGBT and diode modules under sinusoidal PWM modulation are:

$$P_{cond,IGBT} = 0.5 \left( V_{ceo} \frac{I_{s,pk}}{\pi} + R_{on} \frac{I_{s,pk}^2}{4} \right) + m \cos\varphi \left( V_{ceo} \frac{I_{s,pk}}{8} + R_{on} \frac{I_{s,pk}^2}{3\pi} \right) \quad (6)$$

$$P_{cond,diode} = 0.5 \left( V_{Do} \frac{I_{s,pk}}{\pi} + R_D \frac{I_{s,pk}^2}{4} \right) - m \cos\varphi \left( V_{Do} \frac{I_{s,pk}}{8} + R_D \frac{I_{s,pk}^2}{3\pi} \right) \quad (7)$$

, and the switching losses are also functions of switching frequency $f_{sw}$:

$$P_{sw,IGBT} = \frac{[E_{on}(I_{s,pk}) + E_{off}(I_{s,pk})] f_{sw} V_{dc}}{\pi V_{nom}} \quad (8)$$

$$P_{rec,diode} = \frac{E_{rec}(I_{s,pk}) f_{sw} V_{dc}}{\pi V_{nom}} \quad (9)$$

The power module used in each topology and their respective parameters are specified in Table 1.

TABLE 1: POWER ELECTRONIC MODULE PARAMETERS [9, 10, 11]

| Description (Utilization) | Parameter Symbol | FS400R07A3E3_H6 (Dual inverter) | FS400R12A2T4 (Conventional inverter) | FF450R12KT4P (Boost converter) |
|---|---|---|---|---|
| Collector-emitter voltage | $V_{ces}$ | 705 V | 1200 V | 1200 V |
| Nominal voltage | $V_{nom}$ | 300 V | 500 V | 600 V |
| Nominal current | $I_{nom}$ | 400 A | 300 A | 450 A |
| IGBT turn-on voltage | $V_{ceo}$ | 0.798 V | 0.889 V | 0.78 V |
| Diode turn-on voltage | $V_{Do}$ | 0.95 V | 0.92 V | 0.8 V |
| IGBT on resistance | $R_{on}$ | 2.2 mΩ | 3 mΩ | 2.78 mΩ |
| Diode on resistance | $R_D$ | 1.4 mΩ | 1.78 mΩ | 1.27 mΩ |
| Nominal IGBT turn-on loss | $E_{on}$ | 2.24 mJ | 16.5 mJ | 13.689 mJ |
| Nominal IGBT turn-off loss | $E_{off}$ | 8.165 mJ | 18.272 mJ | 18.31 mJ |
| Nominal diode recovery loss | $E_{rec}$ | 5.151 mJ | 14.331 mJ | 23.936 mJ |

The drivetrain power losses of each topology are given below:

$$P_{loss,inv,dual} = P_{inverter,FC} + P_{inverter,Batt} + P_{cond,motor} \quad (10)$$

$$P_{loss,inv,boost} = P_{inverter} + P_{boost} + P_{cond,motor} \quad (11)$$

Where

$$P_{inverter} = 6 * \left( P_{sw,IGBT} + P_{cond,IGBT} + P_{rec,diode} + P_{cond,diode} \right) \quad (12)$$

$$P_{boost} = P_{sw,boostIGBT} + P_{cond,boostIGBT} + P_{rec,boostdiode} + P_{cond,boostdiode} + I_{FC}^2 R_{ind} \quad (13)$$

$$P_{cond,motor} = 3 I_s^2 R_s \quad (14)$$

3. **LOSS COMPARISON**

A. Drivetrain System Parameters

The fuel cell model utilized in this study is Ballard FCMove fuel 70kW rated PEM fuel cell stack [12]. The fuel cell voltage and power are modelled as functions of current (Fig. 4).

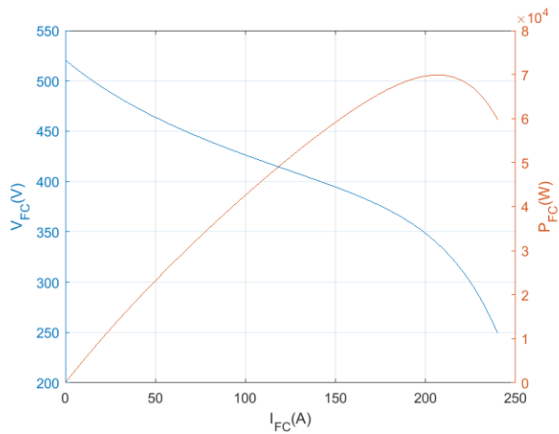

Fig. 4. Ballard FCMove 70kW PEM fuel cell power characteristic

The parameters used in synchronous motor and boost converter modelling are listed in Table 2. In this study, a non-salient motor model was used, where $L_d = L_q = L_s$. The boost converter inductance was calculated to give 10% ripple for a switching frequency of 20 kHz. Likewise, a switching frequency of 20 kHz was selected for the traction inverter in the conventional case.

An advantage of the dual inverter compared to a conventional two-level inverter is its ability to produce motor phase voltage waveforms with a higher number of levels. This increased number of voltage levels reduces the motor phase current ripple. In this paper, it is desired to keep the motor phase current ripple constant between the two cases studied, since this will mean that motor iron losses will also be the same. This condition is desired because iron losses cannot easily be estimated without detailed parameters, which were not available for the motor. As a result of these considerations, both inverters in the dual inverter case are switched at a frequency of 10 kHz.

TABLE 2: SYNCHRONOUS MOTOR AND BOOST CONVERTER PARAMETERS

| Description | Parameter Symbol | Value |
|---|---|---|
| Motor pole pairs | $p$ | 5 |
| Motor synchronous inductance | $L_s$ | 0.838 mH |
| Motor stator resistance | $R_s$ | 45 mΩ |
| Motor magnetic flux linkage | $\varphi_m$ | 0.127 Wb |
| Nominal battery voltage (dual inv) | $V_{batDI}$ | 400 V |
| Nominal battery voltage (boosted) | $V_{batBoost}$ | 800 V |
| Boost converter inductance [13] | $L_{ind}$ | 0.3 mH |
| Boost inductor ESR [13] | $R_{ind}$ | 1.2 mΩ |
| Conventional switching frequency | $f_{swC}$ | 20 kHz |
| Dual inverter switching frequency | $f_{swD}$ | 10 kHz |

B. Simulation of Drivetrain System Operation

A PLECS circuit model of each drive system was used to validate analytical loss calculation method presented in (6-14), and to simulate the impact of converter modulation and higher order harmonics on the obtained loss values. Each model was set to operate the drive system at 50 kW fuel cell output power. Fig. 5 shows a sample of the motor phase voltage and current waveforms of each drive system. The phase voltage was measured from inverter switch node to motor neutral point for the conventional case, and directly across the motor phase for the dual inverter case ($v_{as}$ in Fig. 1).

As discussed in the previous section, the dual inverter case produces a phase voltage waveform with an increased number of levels (9) compared to the conventional case (5). This increase allows a lower switching frequency of 10 kHz to be used for the dual inverter compared to 20 kHz for the conventional inverter, without much sacrificing the phase current quality.

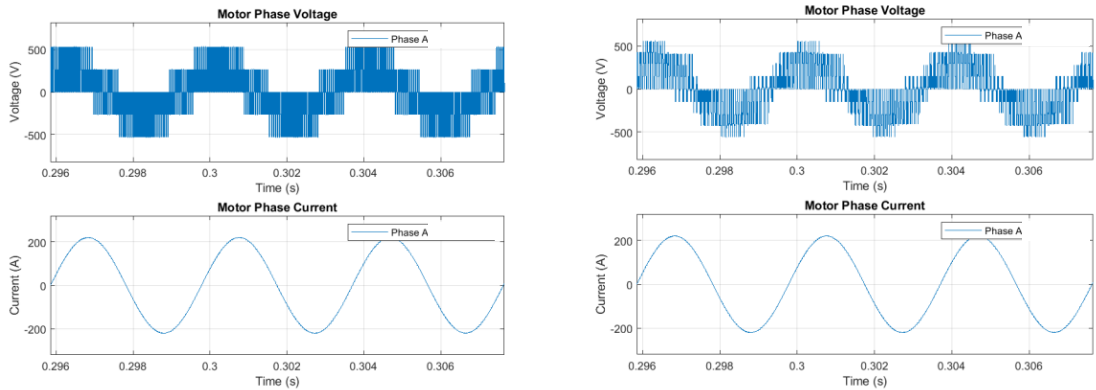

Fig. 5. Simulated motor phase voltage and current waveforms for 50 kW fuel cell power (left: conventional; right: dual inverter)

C. Comparison between simulated and analytical losses

The values obtained from the simulation are compared against the analytical losses in Fig. 6 for a case where 50 kW power was requested from the fuel cell. Only copper losses are considered for the motor and boost inductor. The total losses for the boosted system are 45.93% greater than the total losses for the dual inverter system. A key contributor to this difference is the switching loss of the traction inverter in the boosted case, which is greater than any of the switching losses in the dual inverter configuration due to the higher switching frequency needed (20 kHz vs 10 kHz) and the larger switching energy of the 1.2 kV traction inverter module.

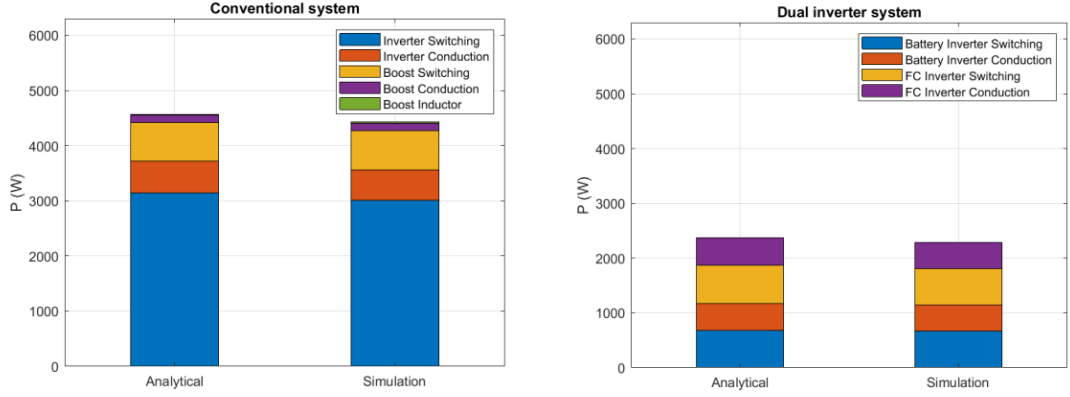

Fig. 6. Analytical vs simulation losses at fuel cell 50kW output

## 4. DRIVE CYCLE ANALYSIS

### A. EPA Driving Cycle

EPA Highway and urban driving cycles (Fig. 7, upper row) are used to assess the energy efficiency of both traction systems. The vehicle parameters utilized in drive cycle analysis are listed in Table 3. These parameters are used to calculate the required drivetrain electrical power $P_{dc}$ from equations (15-18).

Firstly, the required mechanical power from the traction motor of an electric vehicle can be calculated from:

$$P_{ac} = P_{car} + P_{loss,mech} \tag{15}$$

Where $P_{car}$ is the required power for acceleration and $P_{loss,mech}$ is the power lost to air drag and rolling loss. $P_{car}$ is given by:

$$P_{car} = M_{car} v_{car} \frac{dv_{car}}{dt} \tag{16}$$

Where $M_{car}$ is the mass of the vehicle and $v_{car}$ is the velocity. $P_{loss,mech}$ is given by:

$$P_{loss,mech} = v_{car} \left( \frac{1}{2} p_{air} C_d A_f v_{car}^2 + C_r M_{car} g \right) \tag{17}$$

Where $p_{air}$ is the air density at sea level, $g$ is the gravity coefficient, $A_f$ is the vehicles frontal while $C_d$ and $C_r$ are the drag and rolling resistance coefficients respectively. The required drivetrain power $P_{dc}$ can then be obtained from:

$$P_{dc} = P_{ac} + P_{loss,drivetrain} \tag{18}$$

Where $P_{loss,drivetrain}$ is obtained by (10) for the dual inverter case and (11) for the boosted case. During regenerative braking, the inverter loss is subtracted from the regenerated energy flow into the battery packs.

A comparison of drivetrain power losses over both drive cycles for the dual inverter and conventional systems is shown on the lower row of Fig. 7. Clearly, the dual inverter case has significantly lower losses than the conventional case for both drive cycles.

TABLE 3: FORD FOCUS EV VEHICLE PARAMETERS

| Description | Parameter Symbol | Value |
|---|---|---|
| Vehicle mass [14] | $M_{car}$ | 1642.9 kg |
| Front area | $A_f$ | 2.1 m² |
| Coefficient of air drag [15] | $C_d$ | 0.32 |
| Coefficient of rolling friction [15] | $C_r$ | 0.024 |
| Gear ratio |  | 7.82 |
| Tire radius [16] | $r_{tire}$ | 0.3289 m |

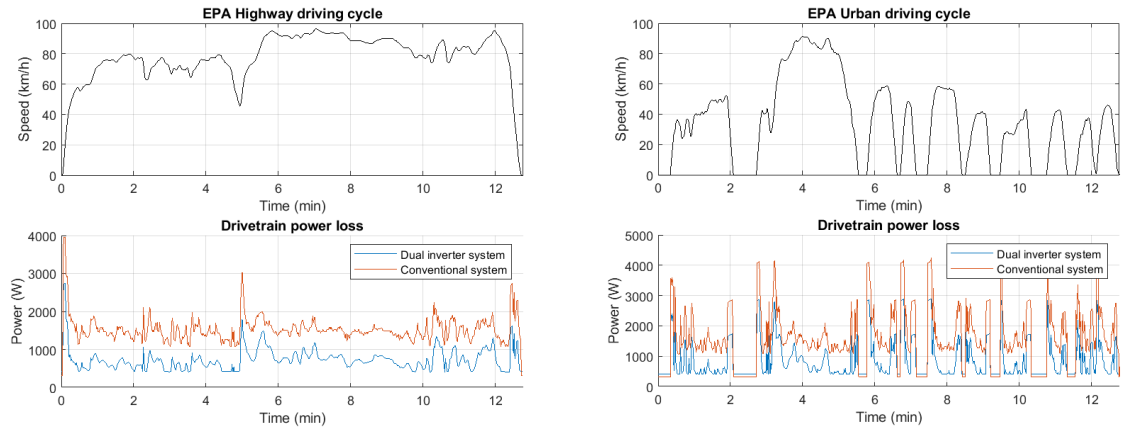

Fig. 7. Power loss of conventional and dual-inverter drivetrain in EPA driving cycles

B. Drivetrain efficiency

For each topology, the energy efficiencies are calculated by:

$$\eta_E = \frac{\int p_{out}(t)dt}{\int p_{out}(t)dt + \int p_{loss,inv}(t)dt + \int p_{loss,motor}(t)dt} \quad (19)$$

where $p_{out}(t)$ equals to $p_{ac}(t)$ while accelerating, and DC-link power output $p_{dc}(t)$ while regenerative breaking. $p_{loss,inv}(t)$ is given by (10) for the dual inverter case and (11) for the boosted case.

The energy efficiency and driving range of each topology for urban and highway settings are summarized in Table 4. The dual inverter case achieves efficiency improvements of 5.27% efficiency improvement during highway driving, and a 10.13% efficiency for urban driving.

TABLE 4: ENERGY EFFICIENCY COMPARISON

| Drive Cycle | Dual Inverter | Conventional |
|---|---|---|
| Highway driving energy efficiency | 94.62% | 89.35% |
| Urban driving energy efficiency | 83.44% | 73.31% |

## 5. DISCUSSION

The dual inverter system achieves an overall higher driving efficiency than the conventional system for both highway and urban driving. This could be attributed to the use of lower voltage rated devices in the dual inverter system. The most important source of loss in the conventional system is the traction inverter switching loss, which is proportional to the switching frequency used and the switching energy of the IGBT modules. The ability of the dual inverter system to use lower voltage rated IGBT modules (with smaller switching energy values) and switch at a lower frequency (due to the multilevel motor phase voltage waveform it produces) causes the switching loss reduction. An overall energy efficiency improvement of 5.27% and 10.13% is achieved for highway and urban driving with the dual inverter compared to the conventional method.